\documentclass{article}
\usepackage{amsmath}
\usepackage{footmisc}
\usepackage{graphicx}
\usepackage{amssymb}
\usepackage{enumerate}
\usepackage{multirow}
\usepackage{longtable}
\usepackage{booktabs}
\usepackage{multicol}
\usepackage{amsthm}
\begin{document}
\cleardoublepage \pagestyle{myheadings}
\bibliographystyle{plain}

\title{A novel type of rogue waves with predictability in nonlinear physics}

\author{Man JIA$^{1}$ and S. Y. Lou$^{1,2}$\thanks{Corresponding author. Email:lousenyue@nbu.edu.cn}\\
\footnotesize{$^1$Physics Department and Ningbo Collaborative Innovation Center}\\
\footnotesize{ of Nonlinear Hazard System of Ocean and Atmosphere, Ningbo University, Ningbo 315211, China}\\
$^2$\footnotesize{Shanghai Key Laboratory of Trustworthy Computing, East China Normal University, Shanghai, 200062, China}}
\vspace{10pt}
\maketitle

\begin{abstract}
Rogue waves named by oceanographers are ubiquitous in nature and appear in a variety of different contexts such as water waves, liquid Helium, nonlinear optics, microwave cavities, etc.
In this letter, we propose a novel type of exact (2+1)-dimensional rogue waves which may be found in many physical fields described by integrable and nonintegrable models.
This type of rogue waves are closely related to  invisible lumps. Usually, a lump is an algebraically localized wave in space but visible at any time. Whence a lump induces a soliton, the lump will become invisible before or after a fixed time. If a bounded twin soliton is induced by a lump, the lump will become a rogue wave (or instanton) and can only be visible at an instant time. 
Because of the existence of the induced visible solitons, the rogue wave may be predictable in some senses. For instance, the height, the position and the arrival time of the rogue wave can be predictable. The idea is illustrated by the celebrate (2+1)-dimensional Kadomtsev-Petviashvili equation in its extended form. 
\end{abstract}

\leftline{\bf PACS numbers: \rm {42.65.Sf, 05.45.-a, 47.27.Sd}}
\vspace{2pc}

Since the 70s of the last century, oceanographers have started to believe the marine folklore for centuries on the rogue waves corresponding to large-amplitude waves surprisingly appearing on the sea surface. Observations gathered by the oil and shipping industries suggest there really is something like a true monster of the deep that devours ships and sailors without mercy or warning.
The study on the rogue wave phenomena has become one of the hot and important subject in natural  science \cite{1}.
Various efforts on the physical mechanisms of the rogue waves have been made by many scientists.
By using the numerical simulations and approximate stability analysis methods, it is pointed out that rogue waves may occur because of the influence of the nonlinear modulation instability \cite{2}. The energy amplification of the rogue waves may come from wave-current interactions. The wave height amplification of rogue waves may also be caused by refraction waves in non-uniform current \cite{3}. Usually, rogue waves appear abruptly and disappear without any trace. Recently, the possibilities to predict the trace of rogue waves had also been done by some scientists. For instance, by using nonlinear time series analysis for some experimental data of three different rogue wave supporting systems, it is found that rogue waves emerge as a consequence of turbulence that means rogue wave events do not necessarily appear without a warning, but are often preceded by a short phase of relative order \cite{4}. Cousins and Sapsis investigate the problem of short-term prediction of rogue waves and have developed and published their research on an effective predictive tool of about 25 wave periods \cite{5}. This tool can give ships and their crews a two-to-three minute warning of potentially catastrophic impact allowing crews some time to shut down essential operations on a ship (or offshore platform).

Theoretically, rational solutions of nonlinear Schr\"odinger (NLS) type systems play a major role in the study of rogue waves for deep water. Rogue waves appear also in many other physical fields where the NLS type systems can be used, especially in nonlinear optics \cite{optical1,optical2}, plasmas \cite{plasma1,plasma2}, atmosphere \cite{fluid1,fluid2}, Bose - Einstein condensations (BECs) \cite{bec} and financial problems \cite{yan}. 

It is also known that rogue waves may also be found in shallow water where the NLS type systems are not valid. To describe shallow water waves there are some integrable systems such as the Korteweg de-Vries (KdV) equation, Kadomtesv-Petviashivili (KP) equation and so on. In our knowledge, the exact pure algebraic rogue waves on these shallow water wave models have not yet been found though the space localized lump solutions for the KPI equation had been known long time ago \cite{Manakov,lump1,lump2}. 

Thus, here, we write down a different type of rogue wave solutions in the form
 \begin{eqnarray}
u&=&\hat{P}(\partial_{x_i}) \ln f,\label{u}\\
f&=&f_0+\sum_{i,j=0}^n  k_{ij}x_ix_j+a\exp\left(\eta\right)
+b\exp\left(-\eta\right),\nonumber \\
&\equiv & f_0+{\xi}^2+a\exp\left(\eta\right)
+b\exp\left(-\eta\right),\ \eta\equiv \sum_{i=1}^nb_{i}x_i,\label{f}
 \end{eqnarray}
where $x_0=1,\ x_n=t$, $\hat{P}$ is an operator function of $\partial_{x_i},\ i=1,\ 2,\ \ldots, n-1,$ $\xi^2=\vec{\xi}\cdot\vec{\xi}=\sum_{m=1}^M\xi_m^2$ is the inner product of the M component vector $\vec{\xi}=\sum_{i=0}^nx_i\vec{K}_i$, 
 $\vec{K}_i=(K_{i1},\ K_{i2},\ \ldots,\ K_{iM}) $ with $n$ and $M$ are positive integers, 
$
k_{ij}=k_{ji}= \vec{K_i}\cdot\vec{K_j}=\sum_{m=1}^M K_{im}K_{jm},\quad i,j=0,\ 1,\ 2,\ \ldots,\ n, $ $a$, $b$ and $K_{im}$ for $i=0,\ 1,\ \ldots,\ n-1,\
 m=1,\ 2,\ \ldots,\ M$  are arbitrary constants
while $f_0,\ b_i$ and $K_{nm}$ for $i=1,\ 2,\ \ldots,\ n,\
 m=1,\ 2,\ \ldots,\ M$ should be model dependent parameters. 
 
 It is clear that the solution (\ref{u}) is a pure algebraic solution for $a=b=0$ which may be a lump (a rogue wave) if it is localized in space only (in both space and time). If $k_{ij}=0$ for all $\{i,\ j\}$, the solution (\ref{u}) is a pure soliton. If $a\neq 0,\ b=0$ the algebraic (lump) part  of (\ref{f}) will be cut at the region $\eta\equiv \sum_{i=1}^nb_{i}x_i>0$ because the term with $a$ will tends to infinity exponentially as $\eta\rightarrow \infty$. Thus, we define the solution (\ref{u}) with (\ref{f}) and $a\neq 0,\ b=0$ (or $b\neq 0,\ a=0$) as ``lumpoff''. For both $a$ and $b$ of (\ref{f}) are nonzero, the lump (algebraic part) will be cut at both sides of $\eta=0$ and then (\ref{u}) becomes a rogue wave for giant amplitude or an instanton for general amplitude.

After finishing some detailed studies, one can find that the solution (\ref{u}) with (\ref{f}) are valid for many integrable and nonintegrable models. For simplicity, here, we take the celebrated KP equation ($n=3,\ x_1\equiv x,\ x_2\equiv y,\ x_3\equiv t$) in its extended form,
\begin{eqnarray}
(u_{t}+u_{xxx}+6 u u_{xx})_x+ \sigma^2 u_{yy}+\gamma (2u^2-uv^2-2vu_x)_{xx}=0,\quad v_x=u, \label{kp}
\end{eqnarray}
as a simple illustration example. The model (\ref{kp}) is called KPI for $\sigma^2+1=\gamma=0$ and KPII for $\sigma^2-1=\gamma=0$. 
The KP equation was firstly derived to study the evolution of long ion-acoustic waves of small amplitude propagating in plasmas under the effect of long transverse perturbations \cite{kp1}. The KP equation was widely accepted as a natural extension of the classical KdV equation to two spatial dimensions, and was later derived as a model for surface and internal water waves \cite{kp2}, in nonlinear optics \cite{kp3} and almost in all other physical fields such as in shallow water waves, ion-acoustic waves in plasmas, ferromagnetics, Bose-Einstein condensation and string theory. The extended KP equation (\ref{kp}) may not be integrable under the sense that its bilinear form can not be cast into the Hirota's form. 

For the extended KP equation (\ref{kp}), the solution (\ref{u}) with (\ref{f}) becomes 
\begin{equation}
u=2(\ln f)_{xx},\ f= f_0+\xi^2+a\ e^{\eta_0}
+b\ e^{-\eta_0},\ \eta_0=k_0x+p_0y+\omega_0t,\label{fkp}
 \end{equation}
where $\xi^2=\vec{\xi}\cdot \vec{\xi}\equiv \sum_{m=1}^M\xi_m^2$,\ $\vec{\xi}$ is an M component vector related to the constant vectors $\vec{k},\ \vec{p},\ \vec{\omega}$ and $\vec{\alpha}$ by 
\begin{equation*}
\vec{\xi}\equiv x\vec{k}+y\vec{p}+t\vec{\omega}+\vec{\alpha}. 
\end{equation*}
The constant vector $\vec{\omega}$ should satisfy the dispersion relation ($k_{12}=\vec{k}\cdot\vec{p}=\sum_{m=1}^Mk_mp_m,\ $ $ k_{11}=k^2=\vec{k}\cdot\vec{k},\ $ $ k_{22}=p^2=\vec{p}\cdot\vec{p} $), 
\begin{equation}
\vec{\omega}=\frac{\sigma^2}{k^2}\left(p^2 \vec{k}-2k_{12}\vec{p}\right), 
 \label{dp}
\end{equation}
while the scalar constants $k_0,\ p_0,\ \omega_0$ and $f_0$ are completely 
determined by the constant vectors $\vec{k}$, $\vec{p}$ and $\vec{\alpha}$,
\begin{equation}
k_0^2=\frac{\sigma^2}{(4\gamma+3)k^4}\left[k_{12}^2-k^2p^2
\right],\ p_0=\frac{k_0k_{12}}{k^2},\ \omega_0=-k_0^3-\sigma^2\frac{p_0^2}{k_0},\label{k0p0}
\end{equation}
and ($k_{02}=\vec{p}\cdot\vec{\alpha},\ k_{01}=\vec{k}\cdot\vec{\alpha},\ k_{00}=\alpha^2=\vec{\alpha}\cdot\vec{\alpha}$)
\begin{equation*}
f_0=\frac{k^2k_{02}^2
+p^2k_{01}^2-2k_{01}
k_{02}k_{12}}{k^2p^2
-k_{12}^2}-\alpha^2+\frac{k^2}{k_0^2}+\frac{abk_0^2}{k^2}.
\end{equation*}

For the KPI equation, the pure line solitons have not yet been found. From the relations (\ref{k0p0}), we know that the possible appearance of the twin-soliton parts ($a$ and $b$ related parts of (\ref{fkp})) are completely induced by the lump parts (algebraic parts of (\ref{fkp}), i.e., $a=b=0$ in (\ref{fkp})).

In the solution (\ref{fkp}), in addition to the arbitrary constants $a$ and $b$,
there are three arbitrary $M$ component constant vectors $\vec{k},\ \vec{p}$ and $\vec{\alpha}$. For the extended KP equation (\ref{kp}), $M$ can be fixed as 3 without loss of generality by redefinition of free  constants. Some special cases of (\ref{fkp}) with $M=2,\ \sigma^2=-1, \ \gamma=0$ and $a^2+b^2\neq 0$ are firstly given by Zhang and Chen \cite{chen} and lately by Yang and Ma \cite{ma}. Although $M$ in (\ref{fkp}) may be arbitrary integer, the solution  (\ref{fkp}) with $a=b=0$ can be proved only a pure lump solution but not rogue wave (instanton) solution because of the dispersion relation (\ref{dp}).

In the pure lump case ($a=0,\ b=0$) for the integrable KP equation ($\gamma=0$), its special forms for $M=2$ had been found long time ago \cite{Manakov} by many authors via different methods, such as Wronskian formulation, the Casoratian formulation and the Grammian or Pfaffian formulation \cite{lump1,lump2}. It is not difficult to find that the real and analytical condition for the solution  (\ref{fkp}) with $\gamma=0$ is $\sigma^2=-1$. In other words, the solution (\ref{fkp}) with $\gamma=0$ is analytic for the KPI and singular for the KPII.

From the expression  (\ref{fkp}), one can see that the solution is constructed by three parts, the pure algebraic part ($a=b=0$ part), $a$-part (the exponential part with the coefficient $a$) which is dominant part for the area $\eta>0$, and $b$-part (the exponential part with the coefficient $b$) which is dominant part for the area $\eta<0$. Furthermore, from the induced wave number relations (\ref{k0p0}), we know that the $a$-part and $b$-part are completely determined by the first algebraic part. In other words, the $a$- and $b$-parts (soliton parts) are induced by the first part (the algebraic lump wave). If there is no the lump part, then there are no the soliton parts. On the other hand, whence the soliton parts are induced, the lump becomes invisible. In the $\eta>0$ region, the $a$-part of (\ref{fkp}) exponentially tends to infinity, the lump part (algebraic part) and $b$-part can be completely neglected. Thus the lump is invisible for $\eta>0$. Similarly, the existence of the induced $b$-parts makes the lump being invisible when $\eta<0$. Only when the invisible ``lump'' moves to the line $\eta\sim 0$, it may become a visible
giant or small peak (rogue wave or instanton).

From the expression  (\ref{fkp}) and the induced wave number conditions (\ref{k0p0}), we know that the special rogue wave (\ref{fkp}) may be predictable in some senses because of the visible parts ($a$- and $b$-parts) including the enough information ($k^2,\ p^2$ and $k_{12}$) of the invisible lump (algebraic) part. Firstly, the approximate moving path of the invisible lump can be read off from the route of lump without the twin soliton parts ($a=b=0$). That means
the (approximate) moving path of the invisible lump, 
\begin{eqnarray}
x=\frac{k_{12}k_{02}
-p^2k_{01}}
{k^2p^2-k_{12}^2}-\sigma^2\frac{p^2}{k^2}t,\quad y=\frac{k_{12}k_{01}
-k^2k_{02}}
{k^2p^2-k_{12}^2}+2\sigma^2\frac{k_{12}}{k^2}t, \label{xy}
\end{eqnarray}
can be obtained from $u_x=u_y=0$ for $a\rightarrow 0$ and $b\rightarrow 0$.

Secondly, the arrival (appeared) time and place of the rogue wave (the instanton) may be predictable because it is related to the cross point of the route (\ref{xy}) and the centerline ($\xi_0+\frac12 \ln \frac{a}{b}$=0) of the twins. The result tells us that the rogue wave will appear at the time
\begin{eqnarray}
t=-\frac32\frac{\sigma^2k^2
k_{01}}
{p^2k^2-k_{12}^2}-\frac{9k^8k_0(\ln a-\ln b)}{4\left(k^2p^2-k_{12}^2\right)^2}\label{t0}
\end{eqnarray}
and the place
\begin{eqnarray}
x= -\frac{p^2\sigma^2}{k^2}t
-\frac{p^2k_{01}-k_{12}k_{02}}{k^2p^2-k_{12}^2}, \quad
y=2\frac{\sigma^2
k_{12}}{k^2}t
-\frac{k^2k_{02}-k_{12}k_{01}}{k^2p^2-k_{12}^2}. \label{xy0}
\end{eqnarray}

Next, the maximum amplitude of the rogue wave (instanton) can be predicted as
\begin{eqnarray}
u_{\max}=\frac{4\sigma^2k^2\left(p^2k^2
-k_{12}^2\right)}
{\sqrt{ab}\sigma^2
\left(p^2k^2-k_{12}^2\right)-(4\gamma+3)k^6}
\label{um}
\end{eqnarray}
which can be simply obtained by calculating the value of $u$ when the ``lump" arrive at the center of the twin-soliton.

Finally, because of the existence of the visible bounded twin-soliton, the interaction between the invisible lump and visible twin-soliton will lead to a visible deformation of the twin-soliton as the rogue wave coming. 

To see more about the special rogue wave (\ref{fkp}), we plot the related figures by fixing free parameters. 

Fig.1 shows the structure of a pure lump (\ref{fkp}) with the parameter selections
\begin{equation}
a=b=\gamma=0,\ k_1 =-k_2=p_1=p_2 =1,\ k_3 =0.2,\ p_3 =0.3,\ \vec{\alpha}=0.\label{param1}
\end{equation}
Fig.1 (a) is a density  plot of the lump at time $t=0$. The three-dimensional plot of Fig.1 (a) is displayed by Fig.1 (b). Fig. 1 (c) shows contour plot of the lump at different times and the 
correct moving routing as described by the straight line (\ref{xy}), i.e., 
$x=209t/204,\ y=-t/17$.

\input epsf
\begin{figure}
\centering\epsfxsize=7cm\epsfysize=5cm\epsfbox{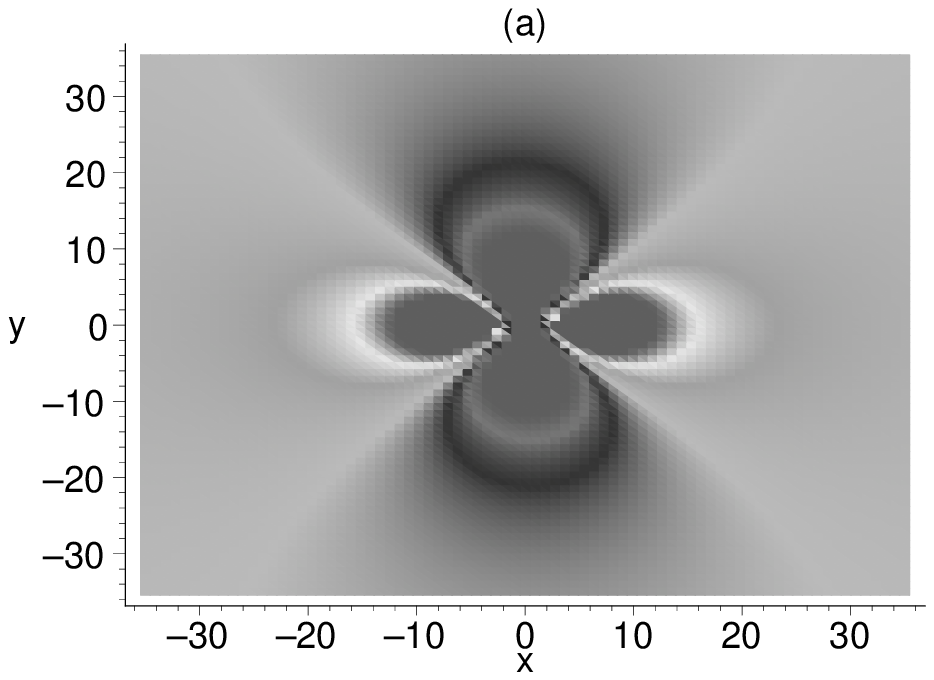}
\centering\epsfxsize=7cm\epsfysize=5cm\epsfbox{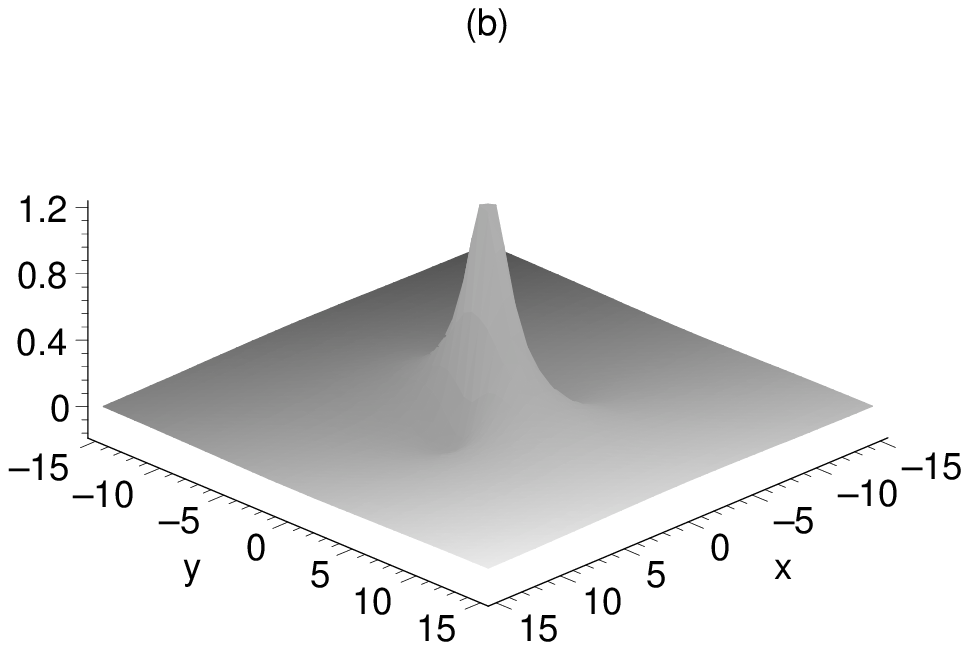}
\centering\epsfxsize=7cm\epsfysize=5cm\epsfbox{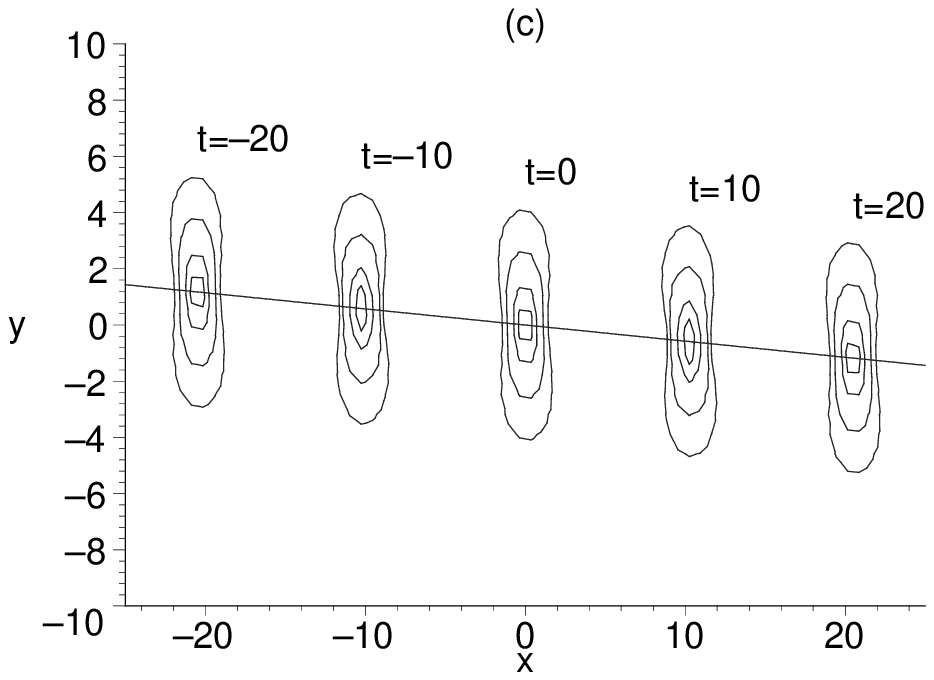}
\caption{Plots of the lump Eq.  (\ref{fkp}) with the parameter selections (\ref{param1}): (a) density plot, (b) three dimensional structure and (c) the contour plot with routing display.}\label{fig1}
\end{figure}

Fig.2 displays the appearance procedure for the lumpoff solution (\ref{fkp}) with the same parameter selections (\ref{param1}) except for $b=0.1$ instead of zero. From Fig.2, we know that the lump is switched off before the lump meets its induced line soliton.

It is also interesting that if we select the same parameter (\ref{param1}) but with $a=0.1,\ b=0$, then the lumpoff solution (\ref{fkp}) will have the disappearance behaviour. The lump will be switched off when it leaves from its induced line soliton. The same figures can be read off from Fig.2 after make the transformation $\{x,\ y,\ t\}\rightarrow \{-x,\ -y,\ -t\}$.       

\input epsf
\begin{figure}
\centering\epsfxsize=7cm\epsfysize=5cm\epsfbox{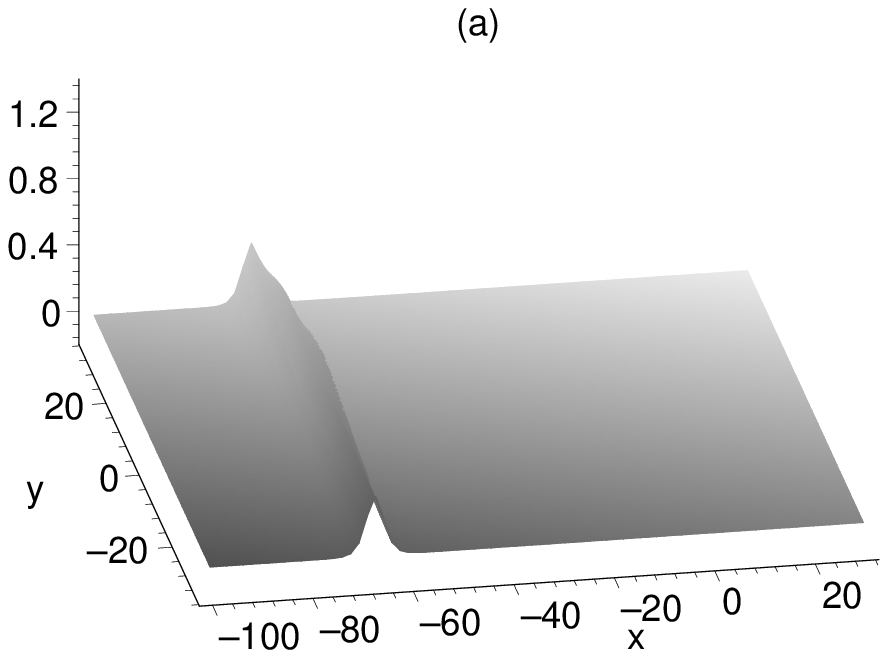}
\centering\epsfxsize=7cm\epsfysize=5cm\epsfbox{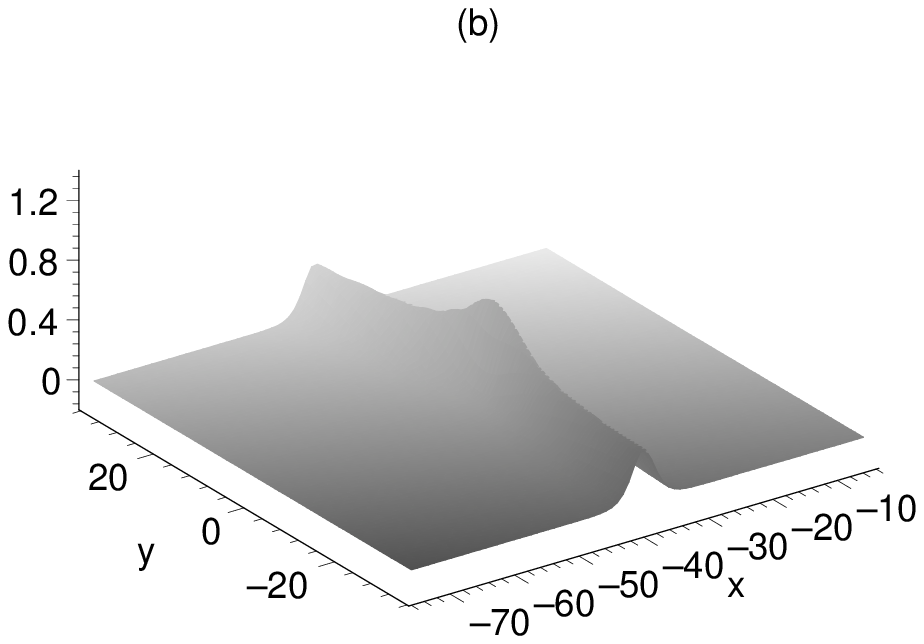}
\centering\epsfxsize=7cm\epsfysize=5cm\epsfbox{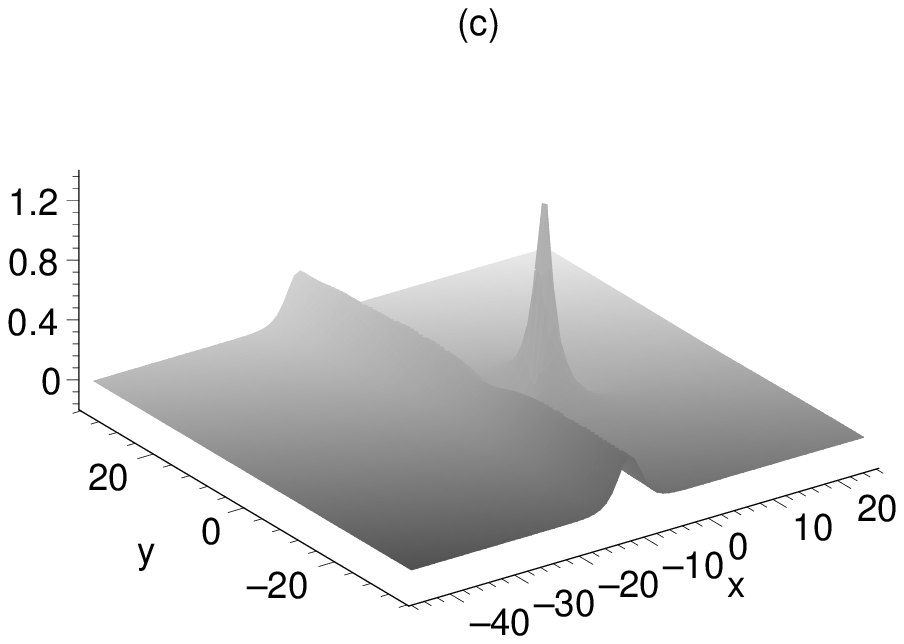}
\centering\epsfxsize=7cm\epsfysize=5cm\epsfbox{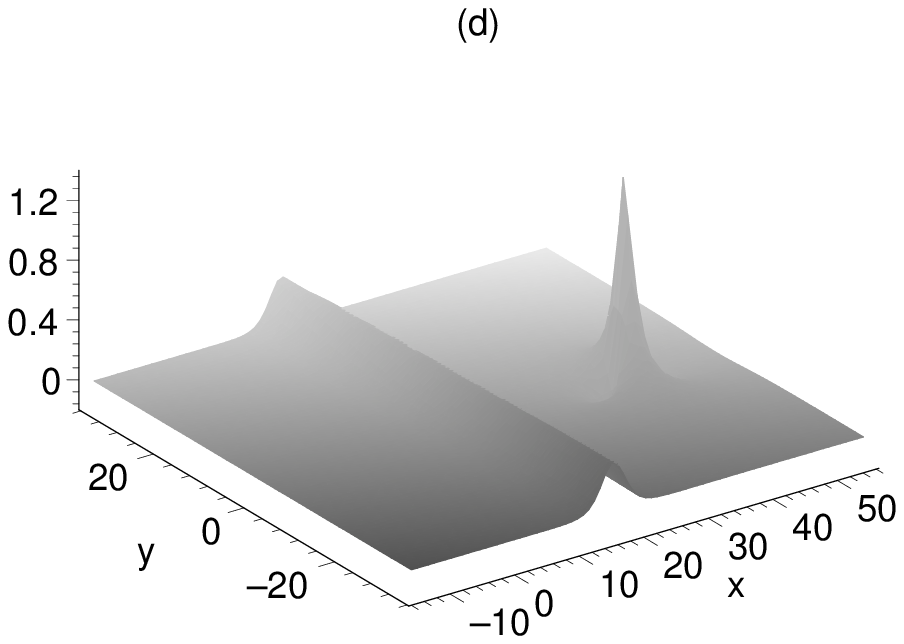}
\caption{The evolution plot of the lumpoff expressed by Eqs. (\ref{fkp}) with the same parameter selections as in (\ref{fig1}) except $b=0.1$ at times (a) $t=-80$, (a) $t=-40$, (c) $t=0$ and (d) $t=40$, respectively.}\label{fig2}
\end{figure}

Figs.3(a)-3(e) exhibit the evolution of the rogue wave solution (\ref{fkp}) with the induced line twin-soliton where the parameters $a=b=0.1$ are used while other parameters are fixed as same as (\ref{param1}).
Fig.3 (f) is a density plot of the rogue wave 
which is appeared at time (\ref{t0}) and space (\ref{xy0}), i.e, $x=y=t=0$ because $\vec{\alpha},\ a$ and $b$ have been selected as $\vec{\alpha}=0$ and $a=b$. 

From Fig. 3, we know that in addition to the 
appeared time and position of the rogue wave, the induced line twin-soliton will be deformed in their shapes and heights as the rogue wave coming. This phenomena will give us a short time warning of potentially catastrophic impact allowing crews some time to shut down essential operations on a ship (or offshore platform). In fact the appearance of the induced twin line soliton with the information (\ref{k0p0}) itself is a warning of potential catastrophic impact.

\input epsf
\begin{figure}
\centering\epsfxsize=7cm\epsfysize=5cm\epsfbox{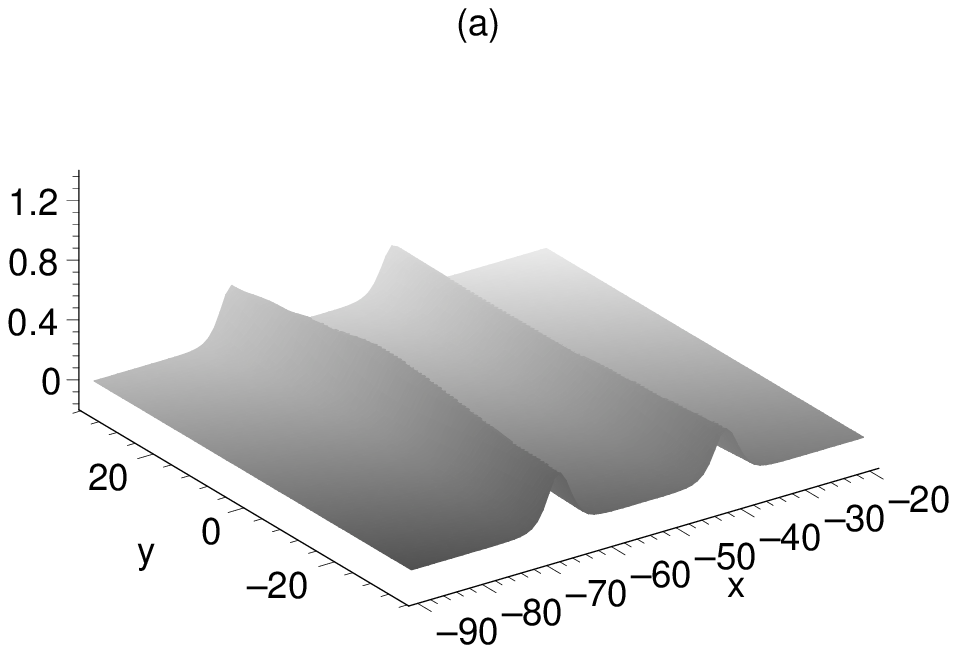}
\centering\epsfxsize=7cm\epsfysize=5cm\epsfbox{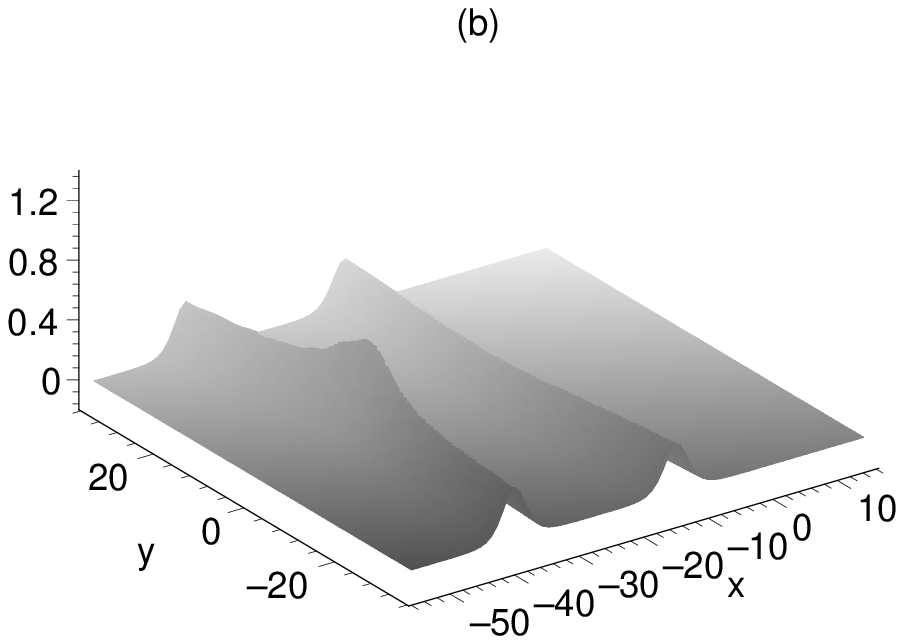}
\centering\epsfxsize=7cm\epsfysize=5cm\epsfbox{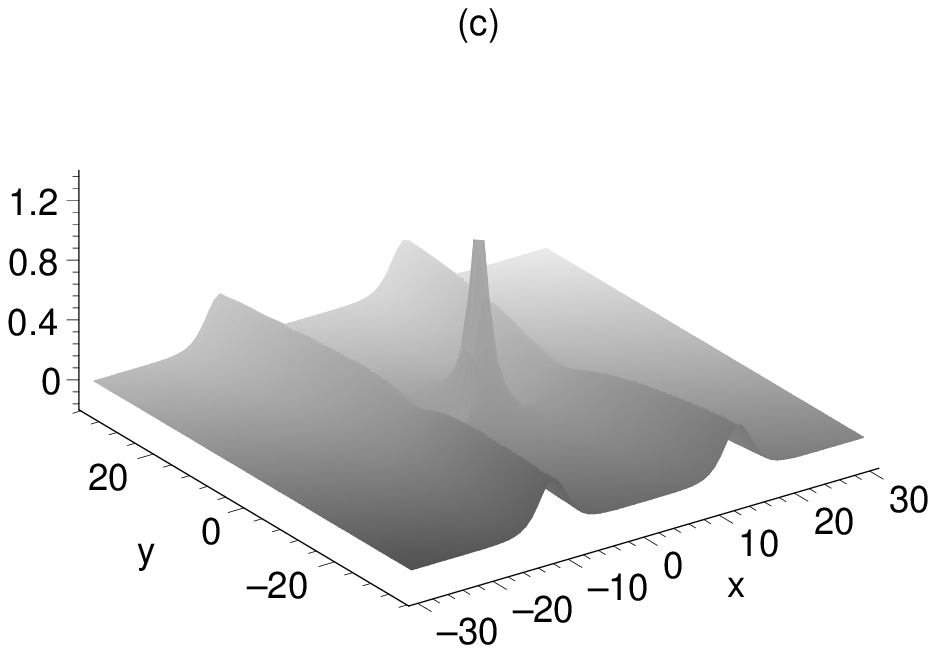}
\centering\epsfxsize=7cm\epsfysize=5cm\epsfbox{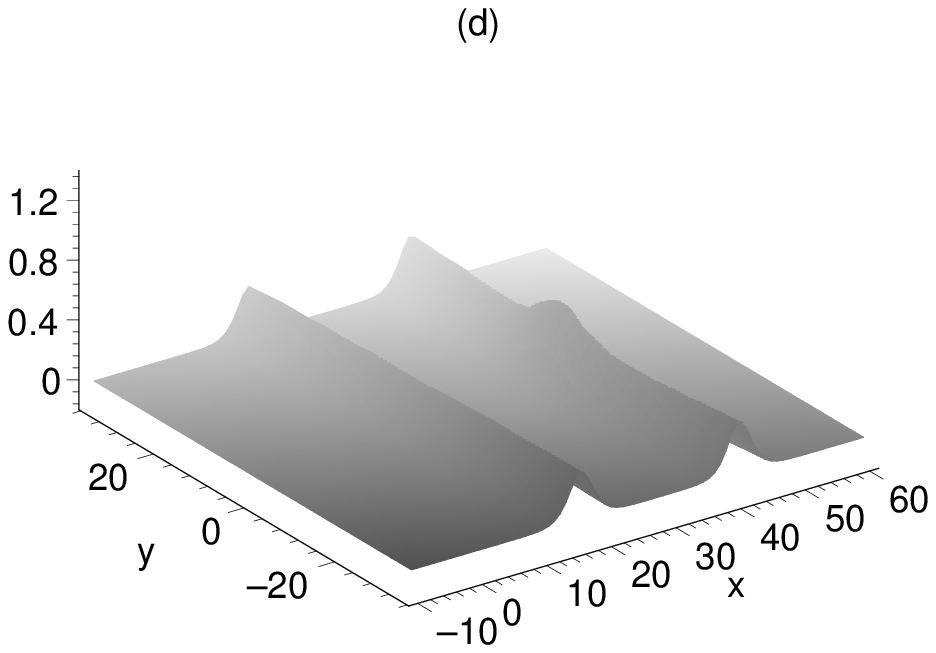}
\centering\epsfxsize=7cm\epsfysize=5cm\epsfbox{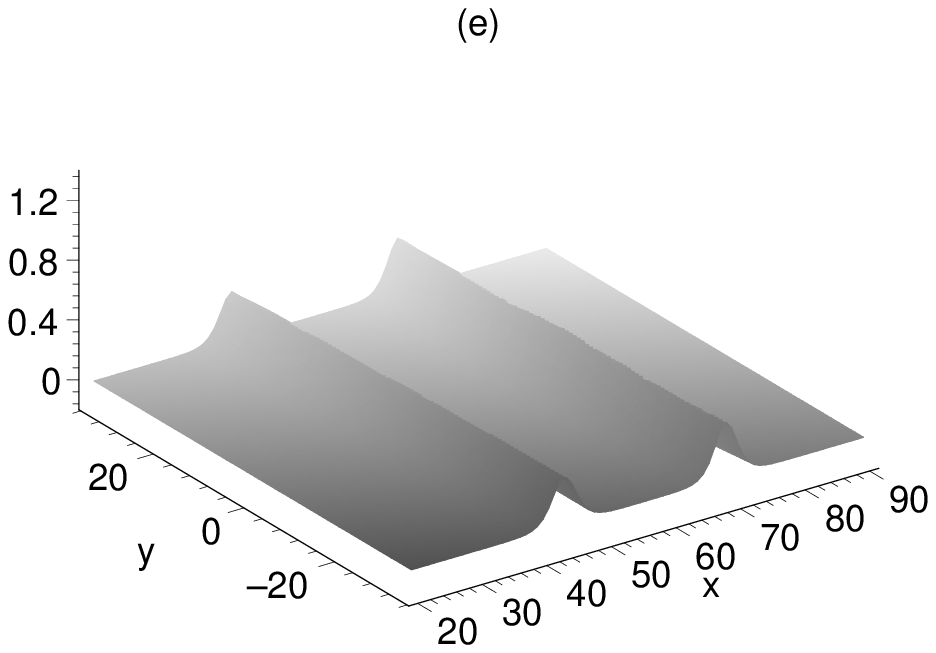}
\centering\epsfxsize=7cm\epsfysize=5cm\epsfbox{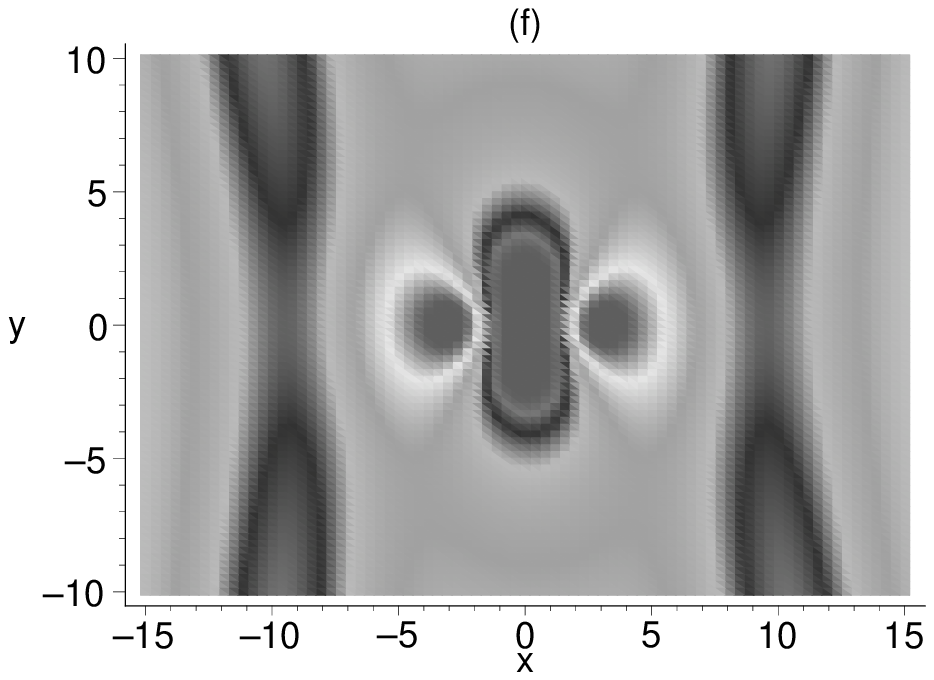}
\caption{The evolution plot of the rogue wave expressed by Eq. (\ref{fkp}) with the same parameter selections as in (\ref{fig1}) but $a=b=0.1$ at times: (a) $t=-80$, (b) $t=-40$, (c) $t=0$, (d) $t=40$ and (e) $t=80$, respectiely. (f) Density plot of (c).  }\label{fig3}
\end{figure}

In summary, a quite general united form (\ref{fkp}) for a lump, lumpoff and rogue wave (instanton) solutions is proposed. In this new type of solution, the dominant part is the lump solution. The lump may induce a single line soliton and/or a line twin-soliton. Whence a single line soliton (characterised by $\eta=0$) is induced by the lump, the lump will be switched off at the area $\eta>0$ ($a\neq 0,\ b=0$) or at the region $\eta<0$ ($b\neq 0,\ a=0$). When the lump is switched off at a half area, it is called lumpoff. When a line twin-soliton is induced, the lump is switched off anywhere  and any time except at a fixed time and fixed point position. In other words the lump becomes a rogue wave for a large amplitude or more generally an instanton for general amplitudes. 

The novel  solution (\ref{u}) is explicitly illustrated by the well known KP equation in its extended form (\ref{kp}). Even for the usual KPI equation ($\sigma^2+1=\gamma=0$ in (\ref{kp})), our result is a generalization of the known results by including more free parameters. Although, the details on the solution (\ref{u}) are discussed only for the extended KP equation. Various integrable and nonintegrable $n+1$ dimensional systems ($n\geq2$) such as the extended Sawada-Kortera, Jimbo-Miwa and Hietarinta systems possess the similar solution structures\cite{Lou,Xia}. 

It should be emphasized that the line soliton parts of (\ref{u}) (exponential parts in (\ref{kp})) are induced from the algebraic (lump) part. If there is no lump part, then there is no such kind of soliton parts with the ``dispersion relation" (\ref{kp}). Whence the solitons are induced, the lump will be cut off to a lumpoff by its induced single line soliton or be compressed to a rogue wave (instanton) by its  induced twin line soliton. Because of the existence of the induced twin-soliton, this kind of rogue wave can be predictable for its appeared time, point position and its maximum amplitude. The induced visible twin line solitons will have small deformations as the rogue wave approaching to the twin. 
Because this kind of solutions is valid for the well known KP equation and various other systems, one may find this kind of rogue waves in many natural fields and reproduced in labs, say, in fluids experiments.   
 
\section*{Acknowledgements}
The authors are indebt to thank Prof. Y. Chen and B. Li for their helpful discussions. The work was supported by NNSFC (Nos. 11675084, 11435005), Ningbo Natural Science Foundation (No. 2015A610159) and granted by the Opening Project of Zhejiang Provincial Top Key Discipline of Physics Sciences in Ningbo University (No. xkzwl1502).  And the authors were sponsored by K. C. Wong Magna Fund in Ningbo University.

\bibliography{apssamp}
\bibliographystyle{elsarticle-num}

\end{document}